\documentclass[aps,pra,floatfix,onecolumn,amsmath,amssymb,superscriptaddress]{revtex4}
\usepackage{times}
\usepackage{latexsym}
\usepackage{graphicx}
\usepackage{verbatim,times,bbm}
\usepackage{color}
\usepackage{appendix}
\usepackage{array}
\usepackage{float}

\usepackage{color}

\begin{document}

\title{Universal feedback control of two-qubit entanglement}

\author{Morteza Rafiee}
\email{m.rafiee178@gmail.com}
\affiliation{Department of Physics, Shahrood University of Technology , 3619995161 Shahrood, Iran}

\author{Alireza Nourmandipour}
\email{anoormandip@stu.yazd.ac.ir}
\affiliation{Atomic and Molecular Group, Faculty of Physics, Yazd University, Yazd 89195-741, Iran}

\author{Stefano Mancini}
\email{stefano.mancini@unicam.it}
\affiliation{School of Science \& Technology, University of Camerino, I-62032 Camerino, Italy}
\affiliation{INFN-Sezione di Perugia, Via A. Pascoli, I-06123 Perugia, Italy}

\begin{abstract}
We  consider two-qubit undergoing local dissipation and subject to local driving. We then determine the optimal \emph{Markovian} feedback action to preserve initial entanglement as well as to create stationary entanglement with the help of an $XY$ interaction Hamiltonian. Such feedback actions are worked out in a way not depending on the initial two-qubit state, whence called \emph{universal}.   
\end{abstract}


\date{\today}

\maketitle

\section{Introduction}

Quantum entanglement has been recognized in recent decades as a resource for quantum information processing \cite{HHHH2009}. As such it should be controllable. Several efforts have been devoted to control entanglement \cite{ENTCON}. 
Control can takes place with open-loop or
and closed-loop strategies according to the principle of controllers design \cite{Rabitz2009}.
Quite generally closed loop control performs better than open loop control because it involves gathering information about the system state and then according to that actuate a corrective action on its dynamics, but results more difficult to implement \cite{Dong2010}.
A good compromise between these two tensions is probably represented by Markovian feedback \cite{Wise94}, who brings the advantages of closed loop control but is not much difficult to realize.
In fact it rests on an actuation based on the measurement result obtained immediately before, 
hence the name Markovian.
Nevertheless it carries an inherent double optimization, over the measurement and over the actuation
\cite{Zhang17}. 
This makes designing optimal control a daunting task even for Markovian feedback, especially when dealing with composite systems and hence with entanglement control (we refer here to \emph{local} control, i.e. measurement and actuation are both local operations).
The best results (in terms of optimality) have been achieved in the context of Gaussian systems \cite{Mancini2007}. For qubit systems, due to their inherent nonlinearity, the situation is more complicate.
With two qubit, on the one hand, a proof of principle of the effectiveness of Markovian feedback in stabilizing entanglement was given in \cite{Mancini2005}, but it is not optimal. On the other hand, the effectiveness of Markovian feedback in protecting initial entangled states has shown in \cite{RNM16}. This action though optimal was derived in a way depending on the initial state.

Here we generalize these results by determining the optimal Markovian feedback action to preserve initial entanglement as well as to create stationary entanglement with the help of an $XY$ interaction Hamiltonian. Moreover, such feedback actions are worked out in a way not depending on the initial 
two-qubit state, whence referred to as \emph{universal}.

The layout of the paper is as follows. We start by introducing the model feedback action in Sec.\ref{sec:model}.
Then we address the issue of preserving initial entanglement in Sec. \ref{sec:preserving} and subsequently the issue of stabilizing entanglement in Sec. \ref{sec:stabilizing}.
Finally, Sec. \ref{sec:conclu} is for conclusion.
Throughout the paper we will use $\iota$ to denote the imaginary unit.

\section{The Model}\label{sec:model}

Consider a two-qubit system whose dynamics 
is governed by the following master equation
\begin{equation}\label{me0}
\dot{\rho}=-\iota\left[H,\rho\right]+{\cal D}\left[\sigma_1\right]\rho
+{\cal D}\left[\sigma_2\right]\rho,
\end{equation}
where $H$ denotes the Hamiltonian and $\sigma_i,\sigma_i^\dag$ ($i=1,2$) are the lowering, raising Pauli operator. Furthermore
\begin{equation}
{\cal D}\left[c\right]\rho\equiv c\rho c^\dag-\frac{1}{2}c^\dag c\rho-\frac{1}{2}\rho c^\dag c,
\end{equation}
is the dissipative super operator and the way it appears in Eq.\eqref{me0} shows qubit 
dissipation into \emph{local} environments.

Following the reasoning of Ref.\cite{Wise94} and generalizing it, we may think at Eq.\eqref{me0} as coming from averaging selective evolutions under local measurements with probability operator value elements 
\begin{subequations}
\begin{eqnarray}
\Omega_1(dt)&=&\sqrt{dt} \sigma_1,\\
\Omega_2(dt)&=&\sqrt{dt} \sigma_2,\\
\Omega_3(dt)&=&1-\left(\iota H+\frac{1}{2}\sigma_1^\dag \sigma+\sigma_2^\dag\sigma_2\right)dt,
\end{eqnarray}
\end{subequations}
describing detection (jump) on the first (resp. second) environment $\Omega_1$ (resp. $\Omega_2$) and no detection $\Omega_3$.
The measurement time is the infinitesimal $dt$ as it is appropriate for continuous measurement. 
It is then easy to verify that the non selective evolution under this measurement 
\begin{equation}
\rho(t+dt)=\sum_{j=1}^3\Omega_j(dt)\rho(t)\Omega^\dag_j(dt),
\end{equation}
is equivalent to the master equation \eqref{me0}.

The selective evolution allows us to incorporate the feedback action. This one, in order to be Markovian, must cause an immediate state change based only on the result of the measurement in the preceding infinitesimal time interval.
Hence it must occur immediately after a detection and cause a finite amount of evolution.
Let this finite evolution following a detection on qubit $i$ at time $t$ be as 
\begin{equation}
\tilde{\rho}_i(t+dt)=e^{{\cal K}_i}\sigma_i\rho(t)\sigma_i^\dag dt,
\label{calK}
\end{equation}
where ${\cal K}_i$s are Liouville super operators (tilde means that the density operator in unnormalized). Form Eq.\eqref{calK} it is clear that the feedback action is \emph{local}.

The nonselective evolution of the system is then given by 
\begin{equation}
\rho(t+dt)\propto \tilde{\rho}_1(t+dt)+\tilde{\rho}_2(t+dt)+\tilde{\rho}_3(t+dt), 
\end{equation}
where $\tilde{\rho}_3(t+dt)=\Omega_3(dt)\rho(t)\Omega_3^\dag(dt)$.
Since the latter is unchanged by feedback, we get for the normalized density operator
\begin{equation}
\dot\rho=-\iota [H,\rho]+\sum_{i=1}^2\left[e^{{\cal K}_i}\sigma_i\rho\sigma_i^\dag-\frac{1}{2}\sigma_i^\dag\sigma_i\rho-\frac{1}{2}\rho\sigma_i^\dag\sigma_i\right].
\end{equation}
Assuming that ${\cal K}_i$s acting in a Hamiltonian way (so to avoid introducing further noise)
\begin{equation}
{\cal K}_i\rho=-\iota[F_i,\rho],
\end{equation}
we will further get 
\begin{equation}\label{me}
\dot{\rho}=-\iota\left[H,\rho\right]+{\cal D}\left[e^{-\iota F_1}\sigma_1\right]\rho
+{\cal D}\left[e^{-\iota F_2}\sigma_2\right]\rho,
\end{equation}
where $F_1$, $F_2$ are hermitian operators on $\mathbb{C}^2$ to be determined. They play the role of (local) feedback Hamiltonians and they concur to implement 
unitary local actuations $e^{-\iota F_i}$. Hence these latter can be parameterized as follows:
\begin{equation}\label{Fi}
e^{-\iota F_i}=\left(\begin{array}{cc}
e^{-\iota(\alpha_i+\gamma_i)/2}\cos(\beta_i/2) & -e^{-\iota(\alpha_i-\gamma_i)/2}\sin(\beta_i/2) \\
e^{\iota(\alpha_i-\gamma_i)/2}\sin(\beta_i/2) & e^{\iota(\alpha_i+\gamma_i)/2}\cos(\beta_i/2)
\end{array}\right),
\end{equation}
with $0\le \alpha_i,\beta_i,\gamma_i\le 2\pi$ ($i=1,2$).

Furthermore it is 
\begin{equation}
\begin{aligned}
e^{-\iota F_i}\sigma_i&=\left(
\begin{array}{ccc}
-e^{-\iota(\alpha_i-\gamma_i)/2}\sin(\beta_i/2) & & 0\\ \\
e^{\iota(\alpha_i+\gamma_i)/2}\cos(\beta_i/2) & & 0
\end{array}\right) \\
&=-e^{-\iota(\alpha_i-\gamma_i)/2}\sin(\beta_i/2) |1\rangle_i\langle 1| \\
&+e^{\iota(\alpha_i+\gamma_i)/2}\cos(\beta_i/2) |0\rangle_i\langle 1|, 
\end{aligned}
\end{equation}
having assumed $|1\rangle_i$ as the excited state and $|0\rangle_i$ as the ground state of 
the $i$th qubit.

Quite generally we can split the Hamiltonian $H$ into two contributions: a local driving term e.g.
\begin{equation}
H_{drive}=\alpha\sigma_1^{(y)}+\alpha\sigma_2^{(y)},
\end{equation}
with driving amplitude $\alpha\in\mathbb{R}$ and an interaction term $H_{int}$ to be specified.

Then Eq.\eqref{me} explicitly becomes 
\begin{equation}\label{meexp1}
\begin{aligned}
\dot\rho&=-\iota\left[\alpha\sigma_1^{(y)}+\alpha\sigma_2^{(y)},\rho\right]
-\iota\left[H_{int},\rho\right]
\\
&-\sum_{i=1}^{2}\Bigg(\frac{1}{2}\left(|A_i|^2+|B_i|^2\right) \left(|1\rangle_i\langle 1| \rho +\rho |1\rangle_i\langle 1|\right)\\
&+\left(A_i|1\rangle_i\langle 1|+B_i|0\rangle_i\langle 1|\right)\rho\left(A_i^*|1\rangle_i\langle 1|+B_i|1\rangle_i\langle 0|\right)\Bigg),
\end{aligned}
\end{equation}
with
\begin{subequations}
\begin{eqnarray}
A_i&=&-e^{-\iota(\alpha_i-\gamma_i)/2}\sin(\beta_i/2),\\
B_i&=&e^{\iota(\alpha_i+\gamma_i)/2}\cos(\beta_i/2). 
\end{eqnarray}
\end{subequations}
This means that $|A_i|^2+|B_i|^2=1$.
 
At this point the aim would be the optimization of a measure of entanglement over the parameters characterizing $F_1$ and $F_2$, or equivalently $e^{-\iota F_1}$ and $e^{-\iota F_2}$.

The figure of merit we shall employ for entanglement is the concurrence defined as \cite{Wootters98}
\begin{equation}
C(\rho):=\max\left\{0,\lambda_1-\lambda_2-\lambda_3-\lambda_4
\right\},
\label{eq:conc}
\end{equation}
where $\lambda_i$s are, in decreasing order, the non-negative square roots of the moduli of the eigenvalues of 
\begin{equation}
\rho\;
(\sigma_1-\sigma_1^\dag)\otimes (\sigma_2-\sigma_2^\dag) \;\rho^* \;
(\sigma_1-\sigma_1^\dag)\otimes (\sigma_2-\sigma_2^\dag).
\end{equation}
Furthermore, in the following we will distinguish two tasks: entanglement preservation and entanglement stabilization.

\section{Preserving entanglement}\label{sec:preserving}

Suppose we want to preserve as much as possible an initial entangled state
by using feedback and considering $H_{int}=0$ in Eq.\eqref{meexp1}.
Then, in the basis $\{|11\rangle,|10\rangle,|01\rangle,|00\rangle\}$ 
such equation becomes 
\begin{eqnarray}\label{eqmatrix1}
\dot\rho&=&-\iota
\left(\begin{array}{cccc}
0 & -\iota\alpha & -\iota\alpha & 0\\
\iota\alpha & 0 & 0 & -\iota\alpha \\
\iota\alpha & 0 & 0 & -\iota\alpha \\
0 & \iota\alpha & \iota\alpha & 0 
\end{array}\right) \rho
+\iota \rho
\left(\begin{array}{cccc}
0 & -\iota\alpha & -\iota\alpha & 0\\
\iota\alpha & 0 & 0 & -\iota\alpha \\
\iota\alpha & 0 & 0 & -\iota\alpha \\
0 & \iota\alpha & \iota\alpha & 0 
\end{array}\right)
\notag\\
&&+\left(\begin{array}{cccc}
A_1 & 0 & 0 & 0\\
0 & A_1 & 0 & 0 \\
B_1 & 0 & 0 & 0 \\
0 & B_1 & 0 & 0 
\end{array}\right)
\rho
\left(\begin{array}{cccc}
A_1^* & 0 & B_1^* & 0\\
0 & A_1^* & 0 & B_1^* \\
0 & 0 &  & 0 \\
0 & 0 & 0 & 0 
\end{array}\right)\notag\\
&&+\left(\begin{array}{cccc}
A_2 & 0 & 0 & 0\\
B_2 & 0 & 0 & 0 \\
0 & 0 & A_2 & 0 \\
0 & 0 & B_2 & 0 
\end{array}\right)
\rho
\left(\begin{array}{cccc}
A_2^* & B_2^* &  & 0\\
0 & 0 & 0 & 0 \\
0 & 0 & A_2^* & B_2^* \\
0 & 0 & 0 & 0 
\end{array}\right)\notag\\
&& -\frac{1}{2} \left(\begin{array}{cccc}
2 & 0 & 0 & 0\\
0 & 1 & 0 & 0 \\
0 & 0 & 1 & 0 \\
0 & 0 & 0 & 0 
\end{array}\right)\rho
-\frac{1}{2} \rho\left(\begin{array}{cccc}
2 & 0 & 0 & 0\\
0 & 1 & 0 & 0 \\
0 & 0 & 1 & 0 \\
0 & 0 & 0 & 0 
\end{array}\right).
\end{eqnarray}
Writing 
\begin{equation}
\rho=\left(
\begin{array}{cccc}
{\cal A} & {\cal B}_R+\iota {\cal B}_I & {\cal C}_R+\iota {\cal C}_I & {\cal D}_R+\iota {\cal D}_I \\ 
{\cal B}_R-\iota {\cal B}_I & {\cal E} & {\cal F}_R+\iota {\cal F}_I & {\cal G}_R+\iota {\cal G}_I \\ 
{\cal C}_R-\iota {\cal C}_I & {\cal F}_R-\iota {\cal F}_I & {\cal H} & {\cal I}_R+\iota {\cal I}_I \\ 
{\cal D}_R-\iota {\cal D}_I & {\cal G}_R-\iota {\cal G}_I & {\cal I}_R-\iota {\cal I}_I & 1-{\cal A}- {\cal E}-{\cal H}
\end{array}\right),
\label{eq:rhomatrix}
\end{equation}
equation \eqref{eqmatrix1} can be put in the following form:
\begin{equation}\label{eqmatrixnew1}
\dot{\bf v}={\bf M}{\bf v}-{\bf w},
\end{equation}
where ${\bf v}$ is the unknown vector 
\begin{align}
{\bf v}:=  
\begin{array}{ccccccccccccccc} 
( {\cal A}, {\cal B}_R, {\cal B}_I, {\cal C}_R, {\cal C}_I, {\cal D}_R, {\cal D}_I, {\cal E},  {\cal F}_R, {\cal F}_I, {\cal G}_R, {\cal G}_I, {\cal H}, {\cal I}_R, {\cal I}_I ),
\end{array}
\end{align}
with entries depending on time $t$,
while 
 \begin{widetext}
\begin{eqnarray}
&&{\bf M}:=\notag\\ \notag\\
&&\hspace{-0.45cm}\left(\begin{array}{ccccccccccccccc}
-|B_1|^2-|B_2|^2 & -2\alpha & 0 & -2\alpha & 0 & 0 & 0 & 0 & 0 & 0 & 0 & 0 & 0 & 0 & 0\\
{\rm r}_2+\alpha & \chi_1  & 0 & 0 & 0 & -\alpha & 0 & -\alpha & -\alpha & 0 & 0 & 0 & 0 & 0 & 0\\
{\rm i}_2 & 0 & \chi_1 & 0 & 0 & 0 & -\alpha & 0 & 0 & \alpha & 0 & 0 & 0 & 0 & 0\\
{\rm r}_1+\alpha & 0 & 0 &  \chi_2 & 0 & -\alpha & 0 & 0 & -\alpha & 0 & 0 & 0 & -\alpha & 0 & 0 \\
{\rm i}_1 & 0 & 0 & 0 &  \chi_2 & 0 & -\alpha & 0 & 0 & -\alpha & 0 & 0 & 0 & 0 & 0  \\
0 & {\rm r}_1+\alpha &  -{\rm i}_1 & {\rm r}_2+\alpha & -{\rm i}_2 & -1 & 0 & 0 & 0 & 0 & -\alpha & 0 & 0 & -\alpha & 0 \\
0 & {\rm i}_1 & {\rm r}_1+\alpha & {\rm i}_2 & {\rm r}_2+\alpha & 0 & -1 & 0 & 0 & 0 & 0 & -\alpha & 0 & 0 & -\alpha\\
|B_2|^2 & 2\alpha & 0 & 0 & 0 & 0 & 0 & -|B_1|^2 & 0 & 0 & -2\alpha & 0 & 0 & 0 & 0\\
0 & {\rm r}_1+\alpha & {\rm i}_1 &  {\rm r}_2+\alpha &  {\rm i}_2 & 0 & 0 & 0 & -1 & 0 & -\alpha & 0 
& 0 & -\alpha & 0\\
0 &  {\rm i}_1 &  -{\rm r}_1-\alpha & - {\rm i}_2 &  {\rm r}_2+\alpha & 0 & 0 & 0 & 0 & -1 & 0 & -\alpha & 0 & 0 & \alpha\\
\alpha & 0 & 0 & |B_2|^2 & 0 & \alpha & 0 &{\rm r}_1+ 2\alpha & \alpha & 0 & -\frac{1}{2} & 0 & \alpha & 0 & 0\\
0 & 0 & 0 & 0 & |B_2|^2 & 0 & \alpha & {\rm  i}_1 & 0 & \alpha & 0 & -\frac{1}{2} & 0 & 0 & 0\\
|B_1|^2 & 0 & 0 & 2\alpha & 0 & 0 & 0 & 0 & 0 & 0 & 0 & 0 & -|B_2|^2 & -2\alpha & 0\\
\alpha & |B_1|^2 & 0 & 0 & 0 & \alpha & 0 & \alpha & \alpha & 0 & 0 & 0 & {\rm r}_2+2\alpha & -\frac{1}{2} & 0 \\
0 & 0 & |B_1|^2 & 0 & 0 & 0 & \alpha & 0 & 0 & -\alpha & 0 & 0 & {\rm i}_2 & 0 & -\frac{1}{2}         
\end{array}\right),\notag\\
\label{M1}
\end{eqnarray}
 \end{widetext}
with
\begin{subequations}
\begin{eqnarray}
{\rm r}_i&=&\Re\left[A_iB_i^*\right],\\
{\rm i}_i&=&\Im\left[A_iB_i^*\right],\\
\chi_i&=&-\left( \frac{3}{2}- |A_i|^2\right),
\end{eqnarray}
\end{subequations}
and
\begin{equation}
{\bf w}:=\left(
\begin{array}{ccccccccccccccc}
0 & 0 & 0 & 0 & 0 & 0 & 0 & 0 & 0 & 0 & \alpha & 0 & 0 & \alpha & 0
\end{array}\right)^\top.
\end{equation}
Notice that Eq.\eqref{eqmatrixnew1}, thanks to \eqref{M1}, results independent of $\gamma_i$s.  

Let the initial condition be
\begin{equation}
\rho(0)=|\Psi\rangle\langle\Psi|,
\end{equation}
with a generic pure state $|\Psi\rangle$ parametrized as
\begin{equation}
|\Psi\rangle=\cos\theta_3|11\rangle+e^{\iota \phi_2}\cos\theta_1\sin\theta_3\sin\theta_2 |01\rangle
+e^{\iota \phi_1}\cos\theta_2\sin\theta_3 |10\rangle
+e^{\iota \phi_3}\sin\theta_3\sin\theta_2\sin\theta_1 |00\rangle,
\label{psiini}
\end{equation}
being $\theta_i\in[0,\pi/2]$ and $\phi_i\in[0,2\pi)$.
In turn, this means
\begin{equation}
\begin{aligned}
{\cal A}(0)&=\cos^2\theta_3,\\
{\cal B}_R(0)&=\cos\phi_1\cos\theta_3\cos\theta_2\sin\theta_3,\\
{\cal B}_I(0)&=-\sin\phi_1\cos\theta_3\cos\theta_2\sin\theta_3,\\
{\cal C}_R(0)&=\cos\phi_2\cos\theta_3\cos\theta_1\sin\theta_3\sin\theta_2,\\
{\cal C}_I(0)&=-\sin\phi_2\cos\theta_3\cos\theta_1\sin\theta_3\sin\theta_2,\\
{\cal D}_R(0)&=\cos\phi_3\cos\theta_3\sin\theta_3\sin\theta_2\sin\theta_1,\\
{\cal D}_I(0)&=-\sin\phi_3\cos\theta_3\sin\theta_3\sin\theta_2\sin\theta_1,\\
{\cal E}(0)&=\cos^2\theta_2\sin^2\theta_3,\\
{\cal F}_R(0)&=\cos(\phi_1-\phi_2)\cos\theta_2\cos\theta_1\sin^2\theta_3\sin\theta_2,\\
{\cal F}_I(0)&=\sin(\phi_1-\phi_2)\cos\theta_2\cos\theta_1\sin^2\theta_3\sin\theta_2,\\
{\cal G}_R(0)&=\cos(\phi_1-\phi_3)\cos\theta_2\sin^2\theta_3\sin\theta_2\sin\theta_1,,\\
{\cal G}_I(0)&=\sin(\phi_1-\phi_3)\cos\theta_2\sin^2\theta_3\sin\theta_2\sin\theta_1,\\
{\cal H}(0)&=\cos^2\theta_1\sin^2\theta_3\sin^2\theta_2,\\
{\cal I}_R(0)&=\cos(\phi_2-\phi_3)\cos\theta_1\sin^2\theta_3\sin^2\theta_2\sin\theta_1,\\
{\cal I}_I(0)&=\sin(\phi_2-\phi_3)\cos\theta_1\sin^2\theta_3\sin^2\theta_2\sin\theta_1.
\end{aligned}
\label{eq:ini}
\end{equation}
Clearly depending on the values of parameters $\theta_i$ and $\phi_i$ the initial state can be entangled or factorable.
However it is known that randomly picking these parameters entangled states are the most likely \cite{typical}. Hence we will consider the concurrence at a given time $C(\rho(t))$ averaged over all possible initial states and then maximize it over $\alpha_i,\beta_i$.
This will lead to a \emph{universal control} action, i.e. independent of the initial state.
To this end initial states are chosen 
according to the following measure induced by Haar measure on U$(4)$ \cite{ZS01}
\begin{equation}
d\mu(|\Psi\rangle)=\frac{6}{\pi^3}\prod_{i=1}^3\cos\theta_i\left(\sin\theta_i\right)^{(2i-1)}d\theta_id\phi_i.
\end{equation}
Actually it is useful to consider
\begin{equation}
\theta_i:=\arcsin\left(\xi_i^{1/(2i)}\right),
\end{equation}
so to have flat probability densities for $\phi_i\in[0,2\pi]$ and $\xi_i\in[0,1]$
\begin{equation}
\label{Pr}
Pr(\phi_i)=\frac{1}{2\pi}, 
\quad
Pr(\xi_i)=1.
\end{equation}
Eq.\eqref{eqmatrixnew1} is solved analytically (not reported here for the sake of simplicity) subjected to the initial condition \eqref{eq:ini}.
 Then the average concurrence has been calculated numerically at each time $t$ using an ensemble of $10^5$ states.
The procedure is repeated for values of $\alpha_i,\beta_i$ in the range $0,2\pi$ with step $\pi/12$.
Finally, the maximum value is taken as corresponding to optimal feedback and the value  $\alpha_i=\beta_i=0$ is taken as corresponding to no feedback action.

\begin{figure}[H]
	\centering
	\includegraphics[width=0.5\textwidth]{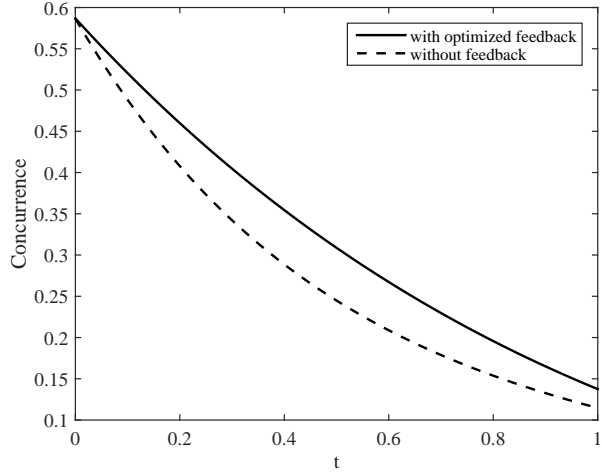}
	\caption{Entanglement measured by concurrence versus time with and without feedback.}
	\label{Cvst}
\end{figure}

The remarkable thing is that the average concurrence does not depend on the driving parameter 
$\alpha$.
Then the results comparing the average concurrence with optimal feedback and without feedback are reported in Fig.(\ref{Cvst}). 
 We can see that feedback is advantageous at any time, although its benefit increases with time, has a maximum at $t=0.4$, and then tends to decrease.
	
Another remarkable result is that the optimal feedback is achieved by the same values of $\alpha_i,\beta_i$ at any time. These values are reported in Tab.\ref{avg_ent} and show a clear asymmetry between the action on the two subsystems.

\begin{table}[h!]
\centering
	\begin{tabular}{|c|c|c|c|} 
		\hline
		$\alpha_1$ & $\beta_1$ & $\alpha_2$ & $\beta_2$   \\ 
		\hline
		$\pi/6$ & $5\pi/6$ & $5\pi/6$ & $\pi$ \\ 
		
		\hline
	\end{tabular}
	\caption{Values of parameters $\alpha_i,\beta_i$ realizing optimal feedback action.}
\label{avg_ent}	
\end{table}

\section{Stabilizing entanglement}\label{sec:stabilizing}

Suppose now we want to stabilize entanglement, i.e. we want to achieve the maximum entanglement at stationary state. 
In this case we also need of an interaction Hamiltonian, e.g.
\begin{equation}
H_{int}=2 J \sigma_1^{(z)}\sigma_2^{(z)}.
\end{equation}
Then we have to solve \eqref{meexp1} with null l.h.s. Subsequently maximize 
the concurrence \eqref{eq:conc} over $\alpha_i$, $\beta_i$, $\gamma_i$, as well as over $\alpha$ and $J$.

In the basis $\{|11\rangle,|10\rangle,|01\rangle,|00\rangle\}$ 
the involved master equation reads
\begin{align}\label{eqmatrix2}
0=&
-\iota
\left(\begin{array}{cccc}
2J & -\iota\alpha & -\iota\alpha & 0\\
\iota\alpha & -2J & 0 & -\iota\alpha \\
\iota\alpha & 0 & -2J & -\iota\alpha \\
0 & \iota\alpha & \iota\alpha & 2J 
\end{array}\right) \rho
+\iota \rho
\left(\begin{array}{cccc}
2J & -\iota\alpha & -\iota\alpha & 0\\
\iota\alpha & -2J & 0 & -\iota\alpha \\
\iota\alpha & 0 & -2J & -\iota\alpha \\
0 & \iota\alpha & \iota\alpha & 2J 
\end{array}\right)
\notag\\
&+\left(\begin{array}{cccc}
A_1 & 0 & 0 & 0\\
0 & A_1 & 0 & 0 \\
B_1 & 0 & 0 & 0 \\
0 & B_1 & 0 & 0 
\end{array}\right)
\rho
\left(\begin{array}{cccc}
A_1^* & 0 & B_1^* & 0\\
0 & A_1^* & 0 & B_1^* \\
0 & 0 &  & 0 \\
0 & 0 & 0 & 0 
\end{array}\right)\notag\\
&+\left(\begin{array}{cccc}
A_2 & 0 & 0 & 0\\
B_2 & 0 & 0 & 0 \\
0 & 0 & A_2 & 0 \\
0 & 0 & B_2 & 0 
\end{array}\right)
\rho
\left(\begin{array}{cccc}
A_2^* & B_2^* &  & 0\\
0 & 0 & 0 & 0 \\
0 & 0 & A_2^* & B_2^* \\
0 & 0 & 0 & 0 
\end{array}\right)\notag\\
& -\frac{1}{2} \left(\begin{array}{cccc}
2 & 0 & 0 & 0\\
0 & 1 & 0 & 0 \\
0 & 0 & 1 & 0 \\
0 & 0 & 0 & 0 
\end{array}\right)\rho
-\frac{1}{2} \rho\left(\begin{array}{cccc}
2 & 0 & 0 & 0\\
0 & 1 & 0 & 0 \\
0 & 0 & 1 & 0 \\
0 & 0 & 0 & 0 
\end{array}\right).
\end{align}
Writing again $\rho$ as \eqref{eq:rhomatrix},
equation \eqref{eqmatrix2} can be put in the same form of \eqref{eqmatrixnew1},
where however now ${\bf M}$ is defined as 
 \begin{widetext}
\begin{eqnarray}
&&{\bf M}:=\notag\\ \notag\\
&&\hspace{-0.45cm}\left(\begin{array}{ccccccccccccccc}
-|B_1|^2-|B_2|^2 & -2\alpha & 0 & -2\alpha & 0 & 0 & 0 & 0 & 0 & 0 & 0 & 0 & 0 & 0 & 0\\
{\rm r}_2+\alpha & \chi_1  & 4J & 0 & 0 & -\alpha & 0 & -\alpha & -\alpha & 0 & 0 & 0 & 0 & 0 & 0\\
{\rm i}_2 & -4J & \chi_1 & 0 & 0 & 0 & -\alpha & 0 & 0 & \alpha & 0 & 0 & 0 & 0 & 0\\
{\rm r}_1+\alpha & 0 & 0 &  \chi_2 & 4J & -\alpha & 0 & 0 & -\alpha & 0 & 0 & 0 & -\alpha & 0 & 0 \\
{\rm i}_1 & 0 & 0 & -4J &  \chi_2 & 0 & -\alpha & 0 & 0 & -\alpha & 0 & 0 & 0 & 0 & 0  \\
0 & {\rm r}_1+\alpha &  -{\rm i}_1 & {\rm r}_2+\alpha & -{\rm i}_2 & -1 & 0 & 0 & 0 & 0 & -\alpha & 0 & 0 & -\alpha & 0 \\
0 & {\rm i}_1 & {\rm r}_1+\alpha & {\rm i}_2 & {\rm r}_2+\alpha & 0 & -1 & 0 & 0 & 0 & 0 & -\alpha & 0 & 0 & -\alpha\\
|B_2|^2 & 2\alpha & 0 & 0 & 0 & 0 & 0 & -|B_1|^2 & 0 & 0 & -2\alpha & 0 & 0 & 0 & 0\\
0 & {\rm r}_1+\alpha & {\rm i}_1 &  {\rm r}_2+\alpha &  {\rm i}_2 & 0 & 0 & 0 & -1 & 0 & -\alpha & 0 
& 0 & -\alpha & 0\\
0 &  {\rm i}_1 &  -{\rm r}_1-\alpha & - {\rm i}_2 &  {\rm r}_2+\alpha & 0 & 0 & 0 & 0 & -1 & 0 & -\alpha & 0 & 0 & \alpha\\
\alpha & 0 & 0 & |B_2|^2 & 0 & \alpha & 0 &{\rm r}_1+ 2\alpha & \alpha & 0 & -\frac{1}{2} & -4J & \alpha & 0 & 0\\
0 & 0 & 0 & 0 & |B_2|^2 & 0 & \alpha & {\rm  i}_1 & 0 & \alpha & 4J & -\frac{1}{2} & 0 & 0 & 0\\
|B_1|^2 & 0 & 0 & 2\alpha & 0 & 0 & 0 & 0 & 0 & 0 & 0 & 0 & -|B_2|^2 & -2\alpha & 0\\
\alpha & |B_1|^2 & 0 & 0 & 0 & \alpha & 0 & \alpha & \alpha & 0 & 0 & 0 & {\rm r}_2+2\alpha & -\frac{1}{2} & -4J \\
0 & 0 & |B_1|^2 & 0 & 0 & 0 & \alpha & 0 & 0 & -\alpha & 0 & 0 & {\rm i}_2 & 4J & -\frac{1}{2}         
\end{array}\right).\notag\\
\end{eqnarray}
 \end{widetext}
Also in this case the dynamics results independent from the $\gamma_i$s. Furthermore we
have no dependence from the initial state, hence the feedback action can again be considered universal.

\begin{figure}[H]
	\centering
	\includegraphics[width=0.5\textwidth]{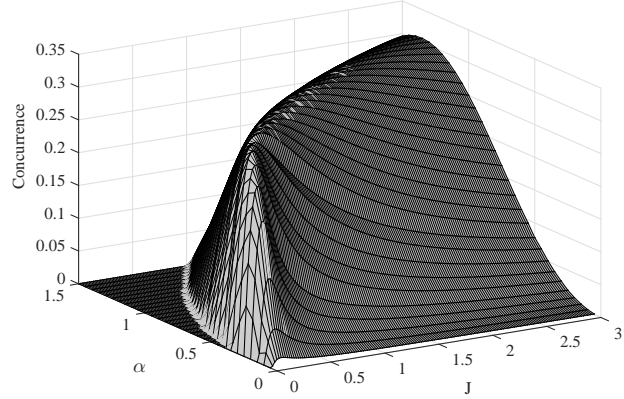}
	\caption{Concurrence achieved with optimal feedback vs $\alpha$ and $J$.\\
	}
	\label{Delta0}
\end{figure}

The solution $\rho(\infty)$ is obtained analytically (not reported here for the sake of simplicity) and then optimization of concurrence has been pursued numerically by varying (for each value of $\alpha$ and $J$)  $\alpha_i$ and $\beta_i$ in the range $0,2\pi$ with step $\pi/12$.
 Finally, the maximum value is taken as corresponding to optimal feedback and the value $\alpha_i = \beta_i = 0$ is taken as corresponding to no feedback action.
The concurrence $C(\rho(\infty))$ achieved with optimal feedback is plotted Fig. \ref{Delta0} vs $\alpha$ and $J$. The results have a mirror symmetry with respect to $\alpha=0$, so only positive values of 
$\alpha$ are considered.

\begin{figure}[h]
	\centering
	\includegraphics[width=0.5\textwidth]{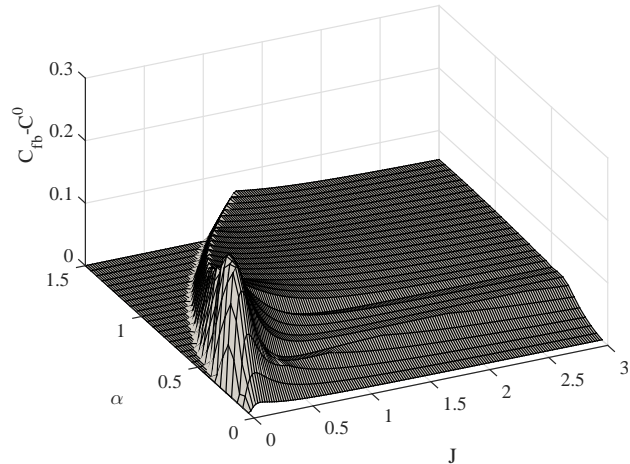}
	\caption{Difference between concurrence achieved with optimal feedback and concurrence achieved without feedback vs $\alpha$ and $J$.\\
	}
\label{Delta1}
\end{figure}

In Fig.\ref{Delta1} is reported the difference between the concurrence
$C(\rho(\infty))$ achieved with optimal feedback and that without feedback action.

\begin{figure}[h]
	\centering
	\includegraphics[width=0.5\textwidth]{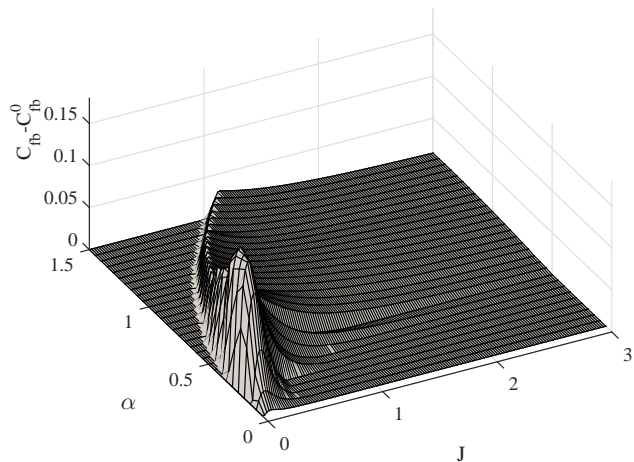}
	\caption{Difference between concurrence achieved with optimal feedback and concurrence achieved with feedback as in Ref.\cite{Mancini2005} vs $\alpha$ and $J$.\\
		}
	\label{Delta2}
\end{figure}

Furthermore, in order to show the supremacy of our optimized feedback, in Fig.\ref{Delta2} we plotted  the difference between the concurrence $C(\rho(\infty))$ achieved with optimal feedback
and that with suboptimal feedback of Ref.\cite{Mancini2005}. 

\begin{figure}[h]
	\centering
	\includegraphics[width=0.5\textwidth]{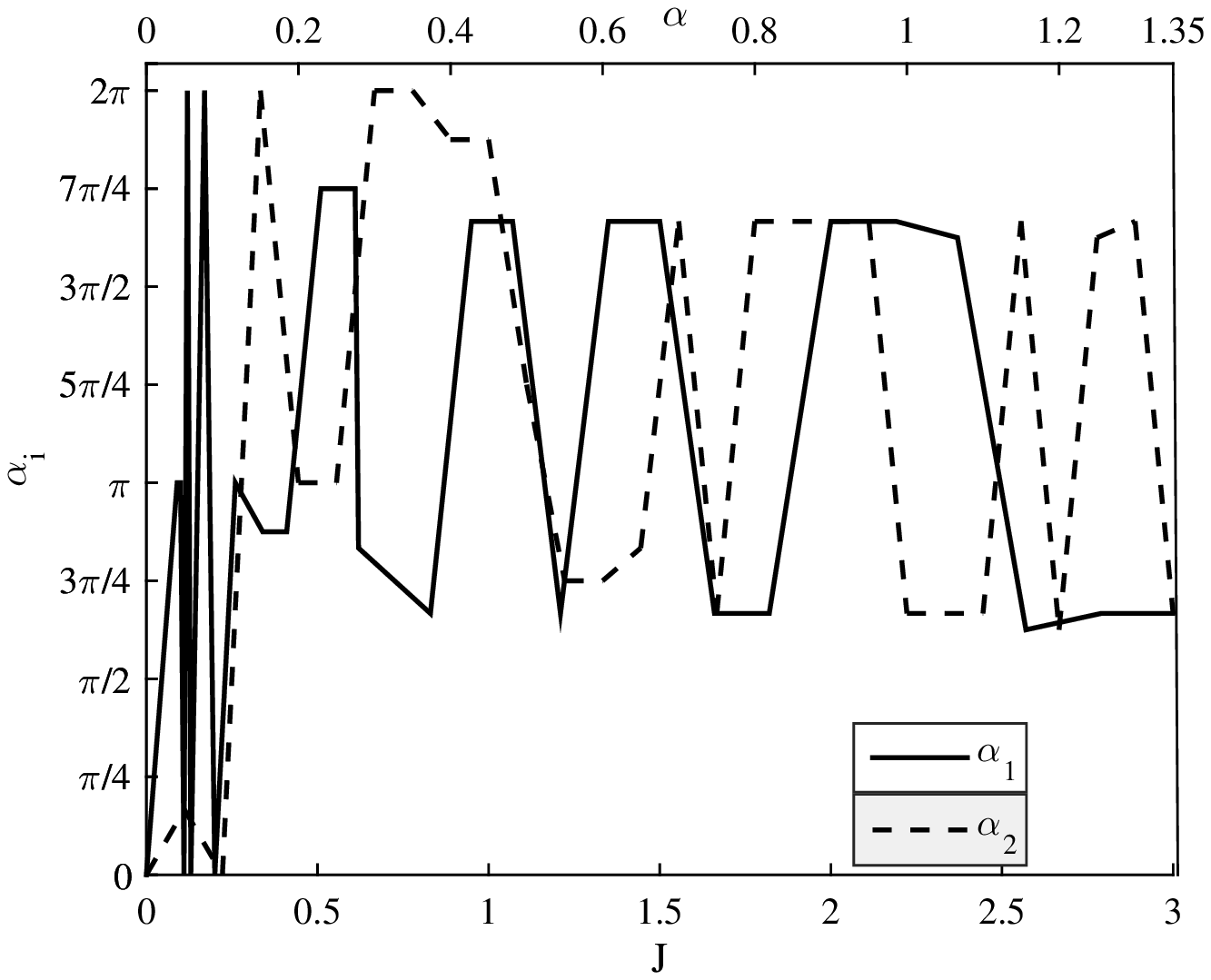}
	\caption{Optimal values of $\alpha_1$ and $\alpha_2$ leading to maximum amount of entanglement along the curve $\alpha=\mathfrak{a}(J)$.\\
	}
	\label{alphai}
\end{figure}

\begin{figure}[h]
	\centering
	\includegraphics[width=0.5\textwidth]{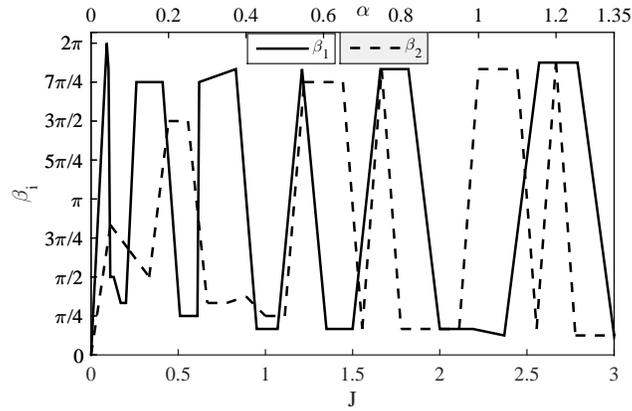}
	\caption{Optimal values of $\beta_1$ and $\beta_2$ leading to maximum amount of entanglement along the curve $\alpha=\mathfrak{a}(J)$.\\
		}
	\label{betai}
\end{figure}

By referring to Fig.\ref{Delta0} we may notice that for each value of $J$ there is an optimal value of $\alpha$ giving the largest concurrence. This determines a curve $\alpha=\mathfrak{a}(J)$ in the plane $J,\alpha$ along which we have maxima of concurrence.
Then, in Figs.\ref{alphai} and \ref{betai} we show the values of parameters $\alpha_i$ and $\beta_i$ respectively that allow to attain the maxima values of concurrence along $\mathfrak{a}$. As we can see they oscillate and depend sensibly to the values of $\alpha$ and $J$.


\section{Conclusion}\label{sec:conclu}

We have addressed two main problems when controlling entanglement
in two-qubit dissipating into their own environments, 
namely protecting initial entanglement and stabilizing entanglement.
We have determined for both tasks optimal Markovian feedback control
by resorting to analytical solutions of the dynamics as well as 
to numerical optimization of concurrence. 
The feedback actions are worked out in a way not depending on the initial two-qubit state, hence 
resulting \emph{universal}.

The present work fill the gap of Ref.  \cite{Mancini2005} where a proof of principle of the effectiveness of Markovian feedback in stabilizing entanglement was given, but it was not optimal, 
as well as of Ref. \cite{RNM16} where the optimal Markovian feedback in protecting initial entangled states was derived in a way depending on the initial state and not continuous in time.

The found results could be helpful in designing experiments
of entanglement control, particularly in settings such as
cavity QED, trapped ions, solid-state based qubit \cite{KCGBG2011}.

The presented analysis can be extended rather easily to other interaction Hamiltonians,
or even to more than two-qubit.
More challenging seems the exploitation of Bayesian
(state-estimation-based) feedback control of two-qubit entanglement, 
following up single-qubit control performed in Ref. \cite{WMW02}.


\end{document}